\pdfoutput=1

\documentclass[12pt,a4paper]{article}

\usepackage{ifthen}
\newboolean{pdflatex}
\setboolean{pdflatex}{true}

\newboolean{articletitles}
\setboolean{articletitles}{true}

\newboolean{uprightparticles}
\setboolean{uprightparticles}{false}

\newboolean{inbibliography}
\setboolean{inbibliography}{false}

\usepackage[top=1in, bottom=1.25in, left=1in, right=1in]{geometry}

\usepackage{microtype}
\usepackage{lineno}
\usepackage{xspace}
\usepackage{caption}

\usepackage{graphicx}
\usepackage{color}
\usepackage{colortbl}
\graphicspath{{./figs/}}
\usepackage{tikz}
\usetikzlibrary{arrows}
\usetikzlibrary{calc}
\usetikzlibrary{chains}
\usetikzlibrary{circuits.logic.IEC,circuits.ee.IEC}
\usetikzlibrary{decorations.pathmorphing}
\usetikzlibrary{decorations.markings}
\usetikzlibrary{fit}
\usetikzlibrary{positioning}
\usetikzlibrary{shadows}
\usetikzlibrary{shapes}
\usetikzlibrary{shapes.geometric}
\usetikzlibrary{through}
\usetikzlibrary{trees}
\usetikzlibrary{mindmap}
\usetikzlibrary{patterns,fadings}
\usetikzlibrary{intersections}
\usetikzlibrary{plotmarks}

\usepackage{amsmath}
\usepackage{amssymb}
\usepackage{amsfonts}
\usepackage{upgreek}
\usepackage{siunitx}
\usepackage{hyperref}
\usepackage[all]{hypcap}

\usepackage{cleveref}
\crefname{chapter}{Ch.\@}{Chs.\@}
\crefname{section}{Sec.\@}{Secs.\@}
\crefname{subsection}{Sec.\@}{Secs.\@}
\crefname{appendix}{Appendix\@}{Appendices\@}
\crefname{figure}{Fig.\@}{Figs.\@}
\crefname{table}{Table\@}{Tables\@}
\crefname{equation}{Eq.\@}{Eqs.\@}

\usepackage{xspace} 
\usepackage{upgreek}

\def\lhcb {\mbox{LHCb}\xspace}

\def\MagUp {\mbox{\em Mag\kern -0.05em Up}\xspace}

\ifthenelse{\boolean{uprightparticles}}%
{

 \def\PDelta      {\ensuremath{\Delta}\xspace}                 
 \def\PXi      {\ensuremath{\Xi}\xspace}                 
 \def\PLambda      {\ensuremath{\Lambda}\xspace}                 
 \def\PSigma      {\ensuremath{\Sigma}\xspace}                 
 \def\POmega      {\ensuremath{\Omega}\xspace}                 
 \def\PUpsilon      {\ensuremath{\Upsilon}\xspace}                 
 
 \def\PB      {\ensuremath{\mathrm{B}}\xspace}                 
                  
 \def\PD      {\ensuremath{\mathrm{D}}\xspace}

 \def\PK      {\ensuremath{\mathrm{K}}\xspace}

 \def\Pi      {\ensuremath{\mathrm{i}}\xspace}

}
{

 \mathchardef\PDelta="7101
 \mathchardef\PXi="7104
 \mathchardef\PLambda="7103
 \mathchardef\PSigma="7106
 \mathchardef\POmega="710A
 \mathchardef\PUpsilon="7107
                  
 \def\PB      {\ensuremath{B}\xspace}                 
                  
 \def\PD      {\ensuremath{D}\xspace}

 \def\PK      {\ensuremath{K}\xspace}

 \def\Pi      {\ensuremath{i}\xspace}

}

\makeatletter
\ifcase \@ptsize \relax
  \newcommand{\miniscule}{\@setfontsize\miniscule{4}{5}}
\or
  \newcommand{\miniscule}{\@setfontsize\miniscule{5}{6}}
\or
  \newcommand{\miniscule}{\@setfontsize\miniscule{5}{6}}
\fi
\makeatother

\DeclareRobustCommand{\optbar}[1]{\shortstack{{\miniscule (\rule[.5ex]{1.25em}{.18mm})}
  \\ [-.7ex] $#1$}}

  \def\Kbar    {{\kern 0.2em\overline{\kern -0.2em \PK}{}}\xspace}

\def\KorKbar    {\kern 0.18em\optbar{\kern -0.18em K}{}\xspace}

  \def\Dbar    {{\kern 0.2em\overline{\kern -0.2em \PD}{}}\xspace}

\def\DorDbar    {\kern 0.18em\optbar{\kern -0.18em D}{}\xspace}

\def\Bbar    {{\ensuremath{\kern 0.18em\overline{\kern -0.18em \PB}{}}}\xspace}

\def\BorBbar    {\kern 0.18em\optbar{\kern -0.18em B}{}\xspace}

\def\Bdb     {{\ensuremath{\Bbar{}^0}}\xspace}

  \def\Y#1S{\ensuremath{\PUpsilon{(#1S)}}\xspace}

\def\Lbar        {{\ensuremath{\kern 0.1em\overline{\kern -0.1em\PLambda}}}\xspace}
\def\LorLbar    {\kern 0.18em\optbar{\kern -0.18em \PLambda}{}\xspace}

\newcommand{\dm}{{\ensuremath{\Delta m}}\xspace}

\def\AT#1     {\ensuremath{A_{\mathrm{T}}^{#1}}\xspace}           

\def\C#1      {\ensuremath{\mathcal{C}_{#1}}\xspace}                       
\def\Cp#1     {\ensuremath{\mathcal{C}_{#1}^{'}}\xspace}                    
\def\Ceff#1   {\ensuremath{\mathcal{C}_{#1}^{\mathrm{(eff)}}}\xspace}        
\def\Cpeff#1  {\ensuremath{\mathcal{C}_{#1}^{'\mathrm{(eff)}}}\xspace}       
\def\Ope#1    {\ensuremath{\mathcal{O}_{#1}}\xspace}                       
\def\Opep#1   {\ensuremath{\mathcal{O}_{#1}^{'}}\xspace}                    

\newcommand{\tev}{\ifthenelse{\boolean{inbibliography}}{\ensuremath{~T\kern -0.05em eV}\xspace}{\ensuremath{\mathrm{\,Te\kern -0.1em V}}}\xspace}
\newcommand{\gev}{\ensuremath{\mathrm{\,Ge\kern -0.1em V}}\xspace}
\newcommand{\mev}{\ensuremath{\mathrm{\,Me\kern -0.1em V}}\xspace}
\newcommand{\kev}{\ensuremath{\mathrm{\,ke\kern -0.1em V}}\xspace}
\newcommand{\ev}{\ensuremath{\mathrm{\,e\kern -0.1em V}}\xspace}
\newcommand{\gevc}{\ensuremath{{\mathrm{\,Ge\kern -0.1em V\!/}c}}\xspace}
\newcommand{\mevc}{\ensuremath{{\mathrm{\,Me\kern -0.1em V\!/}c}}\xspace}
\newcommand{\gevcc}{\ensuremath{{\mathrm{\,Ge\kern -0.1em V\!/}c^2}}\xspace}
\newcommand{\gevgevcccc}{\ensuremath{{\mathrm{\,Ge\kern -0.1em V^2\!/}c^4}}\xspace}
\newcommand{\mevcc}{\ensuremath{{\mathrm{\,Me\kern -0.1em V\!/}c^2}}\xspace}

\def\barn{\ensuremath{\mathrm{ \,b}}\xspace}

\def\ps   {\ensuremath{{\mathrm{ \,ps}}}\xspace}
\def\fs   {\ensuremath{\mathrm{ \,fs}}\xspace}

\def\gsim{{~\raise.15em\hbox{$>$}\kern-.85em
          \lower.35em\hbox{$\sim$}~}\xspace}
\def\lsim{{~\raise.15em\hbox{$<$}\kern-.85em
          \lower.35em\hbox{$\sim$}~}\xspace}

\def\tell1  {TELL1\xspace}
\def\ukl1   {UKL1\xspace}

\DeclareMathAlphabet      {\mathbfit}{OML}{cmm}{b}{it}
\newcommand{\vect}[1]{\ensuremath{\vec{#1}}\xspace}
\newcommand\given[1][]{\:#1\vert\:}

\newcommand{\asymbol}[3]{
\ensuremath{#1
^{\ForEach{;}{
\ifboolexpr{test {\ifnumcomp{\the\thislevelcount}{=}{1}}}
  {\text{\thislevelitem}}
  {, \text{\thislevelitem}}
}{#2}}
_{\ForEach{;}{
\ifboolexpr{test {\ifnumcomp{\the\thislevelcount}{=}{1}}}
  {\text{\thislevelitem}}
  {, \text{\thislevelitem}}
}{#3}}
\xspace}
}

\DeclareSIUnit\clight{\ensuremath{\mathit{c}}}

\DeclareSIUnit[per-mode=symbol]\eVc{\eV\per\clight}
\DeclareSIUnit[per-mode=symbol]\keVc{\kilo\eV\per\clight}
\DeclareSIUnit[per-mode=symbol]\MeVc{\mega\eV\per\clight}
\DeclareSIUnit[per-mode=symbol]\GeVc{\giga\eV\per\clight}
\DeclareSIUnit[per-mode=symbol]\TeVc{\tera\eV\per\clight}

\DeclareSIUnit[per-mode=symbol]\eVcc{\eV\per\square\clight}
\DeclareSIUnit[per-mode=symbol]\keVcc{\kilo\eV\per\square\clight}
\DeclareSIUnit[per-mode=symbol]\MeVcc{\mega\eV\per\square\clight}
\DeclareSIUnit[per-mode=symbol]\GeVcc{\giga\eV\per\square\clight}
\DeclareSIUnit[per-mode=symbol]\TeVcc{\tera\eV\per\square\clight}

\usepackage{cite}
\usepackage{mciteplus}

\begin{document}

\renewcommand{\thefootnote}{\fnsymbol{footnote}}
\setcounter{footnote}{1}


\begin{titlepage}
\pagenumbering{roman}

\vspace*{-1.5cm}
\centerline{\large EUROPEAN ORGANIZATION FOR NUCLEAR RESEARCH (CERN)}
\vspace*{1.5cm}
\noindent
\begin{tabular*}{\linewidth}{lc@{\extracolsep{\fill}}r@{\extracolsep{0pt}}}
\vspace*{-2.7cm}\mbox{\!\!\!\includegraphics[width=.14\textwidth]{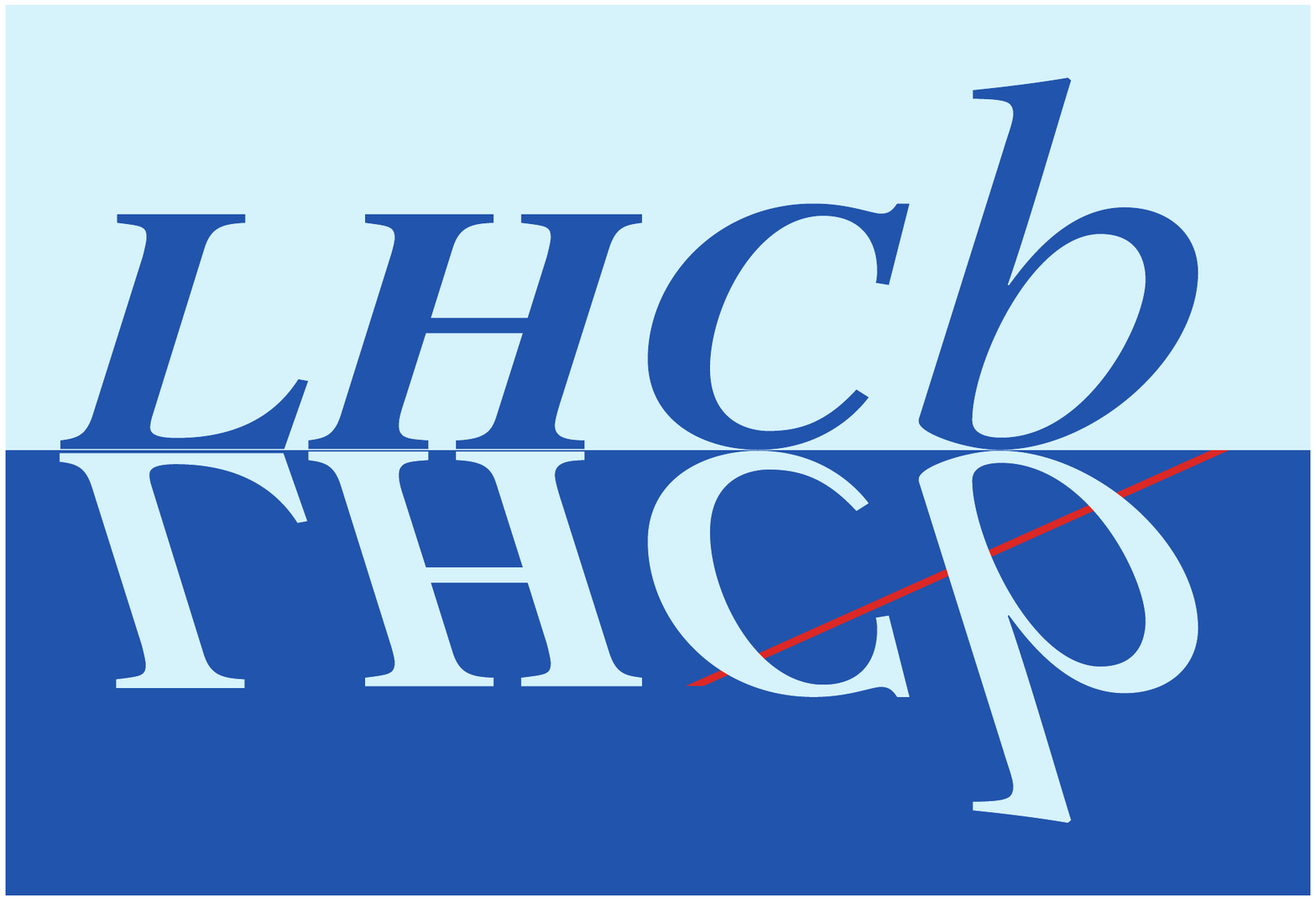}} & & \\
 & & CERN-EP-2016-203 \\  
 & & LHCb-PAPER-2016-037 \\  
 & & January 12, 2017 \\ 
\end{tabular*}

\vspace*{4.0cm}

{\normalfont\bfseries\boldmath\huge
\begin{center}
  Measurement of $C\!P$ violation in $B^0 \!\rightarrow D^+ D^-$ decays
\end{center}
}

\vspace*{2.0cm}

\begin{center}
The LHCb collaboration\footnote{Authors are listed at the end of this Letter.}
\end{center}

\vspace{\fill}

\begin{abstract}
  \noindent
The $C\!P$ violation observables $S$ and $C$ in the decay channel \mbox{$B^0 \!\rightarrow D^+ D^-$} are
determined from a sample of proton-proton collisions at center-of-mass
energies of \SIlist{7;8}{TeV}, collected by the LHCb experiment and
corresponding to an integrated luminosity of $\SI{3}{fb^{-1}}$. The observable $S$
describes $C\!P$ violation in the interference between mixing and the decay amplitude, and
$C$ parametrizes direct $C\!P$ violation in the decay.
The following values are obtained from a flavor-tagged, decay-time-dependent
analysis:
\begin{align*}
  S &=  -0.54 \, ^{+0.17}_{-0.16} \, \text{(stat)} \pm 0.05 \, \text{(syst)}\,, \\
  C &=  \phantom{-}0.26 \, ^{+0.18}_{-0.17} \, \text{(stat)} \pm 0.02 \, \text{(syst)}\,.
\end{align*}
These values provide evidence for $C\!P$ violation at a
significance level of \num{4.0} standard deviations. The phase shift due to
higher-order Standard Model corrections is constrained to a small value of
\begin{align*}
  \Delta\phi = -0.16\,^{+0.19}_{-0.21}\,\si{\radian}\,.
\end{align*}
\end{abstract}

\vspace*{2.0cm}

\begin{center}
  Published in Phys.~Rev.~Lett.~117, 261801 (2016)
\end{center}

\vspace{\fill}

{\footnotesize 
\centerline{\copyright~CERN on behalf of the \lhcb collaboration, licence \href{http://creativecommons.org/licenses/by/4.0/}{CC-BY-4.0}.}}
\vspace*{2mm}

\end{titlepage}

\newpage
\setcounter{page}{2}
\mbox{~}
\cleardoublepage

\renewcommand{\thefootnote}{\arabic{footnote}}
\setcounter{footnote}{0}

\pagestyle{plain}
\setcounter{page}{1}
\pagenumbering{arabic}


Studies of beauty hadron decays into pairs of charm hadrons give access to a
multitude of observables that probe the Cabibbo-Kobayashi-Maskawa (CKM) quark
mixing matrix~\cite{Cabibbo,KobayashiMaskawa} of the Standard Model (SM).
Comparisons of these observables with each other and with similar observables
from beauty hadron decays to charmonia allow higher-order SM
contributions, like loop diagrams, to be separated from effects caused by physics beyond the
SM~\cite{Fleischer1999,Gronau:2008ed,Fleischer2007,Jung:2014jfa,Bel:2015wha}.
For example, under the assumption that flavor symmetries hold to a good approximation, higher-order corrections
in the measurement of $\phi_{{s}}$ in \mbox{$B^0_{{s}} \!\rightarrow D^+_{{s}} D^-_{{s}}$}~\cite{LHCb-PAPER-2014-051} can be
constrained.\footnote{The inclusion of charge-conjugate processes is implied
throughout the Letter, unless otherwise noted.}

In the $B^0$ meson system, $C\!P$ violation in the mixing is negligible, as is the decay
width difference $\mathrm{\Delta}\Gamma$ of the mass
eigenstates~\cite{HFAG}.
In contrast, sizable $C\!P$ violation from the interference
between the direct (unmixed) decay into the $C\!P$-even final state $D^+ D^-$ and the decay to the same final state
after $B^0$--$\kern 0.18em\overline{\kern -0.18em B}{}^0$ mixing, or from the interference of different decay
processes, leads to a decay-time-dependent decay rate of
\begin{equation}\label{eq:simpledecayrates}
\begin{aligned}
  \frac{\mathrm{d}\Gamma(t,d)}{\mathrm{d}t}
  &\propto \mathrm{e}^{-t/\tau} 
    \Big(
      1
      - d\, S \sin{\left(\mathrm{\Delta} m t\right)}
      + d\, C \cos{\left(\mathrm{\Delta} m t\right)}
    \Big)\,,
\end{aligned}
\end{equation}
where $t$ is the proper decay time, $d$ represents the $B^0$ flavor at production and takes a
value of $+1$ for mesons whose initial flavor is $B^0$ and $-1$ for ${\kern 0.18em\overline{\kern -0.18em B}{}^0}$,
$\tau$ is the mean lifetime and $\mathrm{\Delta} m$ is the mass difference between the physical $B^0$ meson
eigenstates. The $C\!P$ observables $S$ and $C$ are related to the $B^0$ mixing
phase $\phi_d$ and a phase shift $\Delta{\phi}$ from the decay amplitudes via
${S}/{\sqrt{1-C^2}} = -\sin(\phi_d + \Delta{\phi})$~\cite{DeBruyn:2014oga}. In
the SM, $\phi_d  = 2\beta$, where $\beta\equiv\arg[-(V_{cd}^{}V_{cb}^{\ast})/(V_{td}^{}V_{tb}^{\ast})]$ is an angle of one of the CKM unitary triangles and $V_{qq^\prime}$ are elements of the CKM matrix. If the \mbox{$B^0 \!\rightarrow D^+ D^-$}
decay amplitude can be described by a dominant tree-level $b\!\rightarrow c\overline{c}d$
transition, the phase shift $\Delta{\phi}$ vanishes and the $C\!P$ observables are given by $C = 0$ and $S = -\sin
\phi_d$. The value of the latter has been measured to be $\sin\phi_d = +0.679 \pm
0.020$~\cite{HFAG} in $b\!\rightarrow c\overline{c}s$ decays such as \mbox{$B^0 \!\rightarrow {J\mskip
-3mu/\mskip -2mu\psi\mskip 2mu} {K^0_{\rm\scriptscriptstyle S}}$}, in which
the contribution from loop processes in the decay can be constrained to high
precision~\cite{Frings:2015eva}. In contrast, previous measurements of the $C\!P$ observables in the decay $B^0 \!\rightarrow D^+ D^-$
by the \mbox{BaBar} and \mbox{Belle} collaborations~\cite{Aubert:2008ah,Rohrken:2012ta} give world average values of
$S =
\num{-0.98\pm0.17}$ and $C = \num{-0.31\pm0.14}$~\cite{HFAG}. The values
are at the edge of the physically allowed region of $S^2+C^2\leq1$, which
leaves room for a large value of $\Delta\phi$.


This Letter reports a measurement of $C\!P$ violation in $B^0 \!\rightarrow D^+ D^-$
decays with the LHCb experiment.
The measurement is based on samples of $pp$ collision data corresponding to
integrated luminosities of $1$ and $\SI{2}{fb^{-1}}$ at center-of-mass energies
of $7$ and $8\,\mathrm{\,Te\kern -0.1em V}$, respectively, recorded by the LHCb experiment. The LHCb detector
is a single-arm forward spectrometer covering the \mbox{pseudorapidity} range
$2<\eta <5$, designed for the study of particles containing $b$ or
{\ensuremath{c}\xspace} quarks, and is described in detail in
Refs.~\cite{Alves:2008zz,LHCb-DP-2014-002}. The online event selection is
performed by a trigger, which consists of a hardware stage, based on information
from the calorimeter and muon systems, followed by a software stage, which
applies a full event reconstruction. Simulated events are produced with the software
described in Refs.~\cite{Sjostrand:2007gs,*Sjostrand:2006za,LHCb-PROC-2010-056,Lange:2001uf,Golonka:2005pn,Allison:2006ve,*Agostinelli:2002hh,LHCb-PROC-2011-006}.


Candidate \mbox{$B^0 \!\rightarrow D^+ D^-$} decays are reconstructed through the subsequent decays \mbox{$D^+ \!\rightarrow K^-\pi^+\pi^+$} and
\mbox{$D^+ \!\rightarrow K^-K^+\pi^+$}, with combinations of two \mbox{$D \!\rightarrow KK\pi$} candidates omitted due to the low branching fraction. The kaon and pion
candidates, which must fulfill loose particle identification (PID) criteria, are required to have transverse momentum $p_{\text{T}}>\SI[per-mode=symbol]{100}{\mega\eV\!\per\clight}$, to have a good track quality and to be inconsistent with
originating from a primary vertex (PV). The three hadron tracks must form a good
common vertex and their combined invariant mass has to be in the range
$\SI[per-mode=symbol]{\pm25}{\mega\eV\!\per\square\clight}$ around the known $D^+$ mass~\cite{PDG2014}. The scalar sum of the
$p_{\text{T}}$ of the three hadrons has to exceed $\SI[per-mode=symbol]{1800}{\mega\eV\!\per\clight}$ and the $D^+$ vertex
has to be significantly displaced from all PVs. Defining $\theta_X$ as the angle
between the momentum vector of a particle $X$ and the displacement vector from
the best-matched PV to the $X$ decay vertex, $\cos\theta_{D^+}$ is required to
be positive.

To suppress contributions from misreconstructed \mbox{$D^+_s \!\rightarrow K^-K^+\pi^+$} decays,
which proceed predominantly through \mbox{$D^+_s \!\rightarrow \phi\pi^+$}, \mbox{$D^+ \!\rightarrow K^-\pi^+\pi^+$} candidates
are rejected if, after assigning the kaon mass hypothesis to the $\pi^+$ with
the higher $p_{\text{T}}$, the invariant mass $m(K^-K^+)$ is within $\SI[per-mode=symbol]{10}{\mega\eV\!\per\square\clight}$
of the known $\phi$ meson mass. Furthermore, if the invariant mass
$m(K^-K^+\pi^+)$ is within $\SI[per-mode=symbol]{25}{\mega\eV\!\per\square\clight}$ of the known $D^+_s$ meson mass, the requirement on the PID
information of the higher-$p_{\text{T}}$ pion to be consistent with the pion hypothesis is tightened.
Similarly, protons can be misidentified as pions, resulting in background
contributions from \mbox{$\Lambda^+_c \!\rightarrow K^-p\pi^+$}. To suppress these processes, the pion candidate with
the higher $p_{\text{T}}$ of \mbox{$D^+ \!\rightarrow K^-\pi^+\pi^+$} is required to be well identified
as a pion if $|m(K^-p\pi^+) −- m_{\Lambda^+_c}| <
\SI[per-mode=symbol]{25}{\mega\eV\!\per\square\clight}$.

Candidate $B^0$ mesons are reconstructed from pairs of oppositely charged
$D^\pm$ candidates that form a common vertex. The scalar sum of the $p_{\text{T}}$ of the
$D^\pm$ mesons must exceed $\SI[per-mode=symbol]{5}{\giga\eV\!\per\clight}$. The decay time significance of each
$D^\pm$ meson, defined as its decay time divided by its estimated uncertainty,
is required to be greater than \num{0}, or greater than \num{3} if one of the $D^\pm$ mesons is reconstructed in the
$K^-K^+\pi^+$ final state. This reduces the contamination of \mbox{$B^0 \!\rightarrow D^-K^-K^+\pi^+$} decays. The $B^0$ candidate is required to have momentum $p >
\SI[per-mode=symbol]{10}{\giga\eV\!\per\clight}$, $\cos\theta_{B^0}>\num{0.999}$, and to not originate from the associated PV. A fit to
the full decay chain, in which the $B^0$ production vertex is constrained to
the position of the associated PV, is performed to determine the reconstructed decay time
$t'$ of the $B^0$ candidate, which differs from the true time $t$. Only
candidates with decay times in the range \SIrange{0.25}{10.25}{\ps} are kept.
The invariant mass $m_{D^+D^-}$ of the $B^0$ candidate is calculated from a
similar fit to the full decay chain, while additionally constraining the invariant masses of $K^-\pi^+\pi^+$
and $K^-K^+\pi^+$ to the known $D^+$ mass, and is required to be in the range
\SIrange[per-mode=symbol]{5150}{5500}{\mega\eV\!\per\square\clight}.

Two boosted decision trees~(BDTs)~\cite{Breiman,AdaBoost}, for $B^0$ final
states with two and three kaons, are used to suppress the combinatorial background. Both
are trained on simulated signal samples and on background samples formed from
$B^0$ candidates at high invariant masses ($>\SI[per-mode=symbol]{5500}{\mega\eV\!\per\square\clight}$), and exploit
observables related to the kinematics of the decay, PID information, and track and vertex
quality. The requirements on the BDT classifier outputs are chosen
to optimize the precision of both $C\!P$ observables, $S$ and
$C$.

\begin{figure}[tb]
\centering
\includegraphics[width=0.48\textwidth]{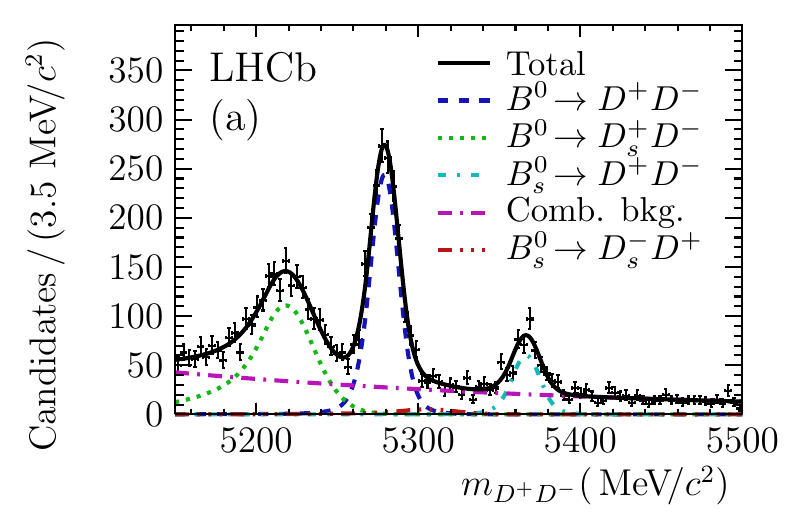}
\includegraphics[width=0.48\textwidth]{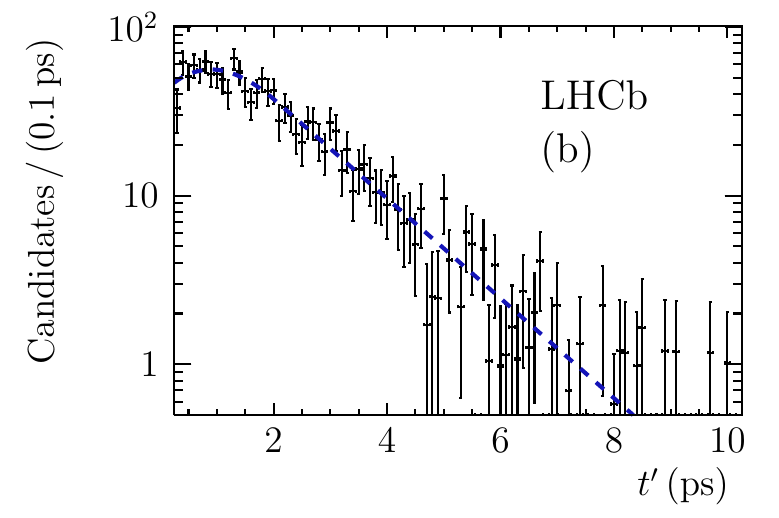}
\caption{Distribution of the reconstructed mass of all \mbox{$B^0 \!\rightarrow D^+
D^-$} candidates (a) and background-subtracted decay time distribution for tagged candidates (b). In
plot (a) besides the data points and the projection of the full PDF (solid
black) the projections of the $B^0$ signal (dashed blue), the \mbox{$B^0_s
\!\rightarrow D^+D^-$} background (short-dash-dotted turquoise), the \mbox{$B^0
\!\rightarrow D^+_s D^-$} background (dotted green), the \mbox{$B^0_s \!\rightarrow
D^-_s D^+$} background (long-dash-three-dotted red) and the combinatorial
background (long-dash-dotted purple) are shown.}
\label{fig:mass_and_decaytime}
\end{figure}

To separate the remaining background from the signal a fit to the $D^+D^-$ invariant mass distribution is performed to calculate signal
candidate weights via the \mbox{\em sPlot} technique~\cite{Pivk:2004ty}. The mass fit is performed simultaneously in four categories, split by the data-taking period (\SI{7}{\tera\eV}, \SI{8}{\tera\eV}) and the number of kaons in the final state. The probability density function (PDF)
used to parametrize the mass distribution consists of four contributions: signal,
\mbox{$B^0_s \!\rightarrow D^+D^-$}, combinatorial background, and a component that includes both \mbox{$B^0 \!\rightarrow D^+_s D^-$}
and \mbox{$B^0_s \!\rightarrow D^-_s D^+$} decays. The signal is modeled by the sum of three Crystal Ball
functions~\cite{Skwarnicki:1986xj} with a common mean. The parameters of the tails
(two towards lower and one towards higher mass) and the three
widths are determined from simulated samples. To account for differences in
the mass resolution in simulation and data, the width parameters are
multiplied by a common scale factor, which is free to vary in the fit to
data. The \mbox{$B^0_s \!\rightarrow D^+D^-$} component shares all shape parameters with the signal PDF
except for the peak position, which is constrained by the known value of the
difference between the $B^0$ and the $B^0_s$ masses~\cite{PDG2014}. Each peak in the \mbox{$B^0 \!\rightarrow D^+_s D^-$} and
\mbox{$B^0_s \!\rightarrow D^-_s D^+$} component is described by the sum of two Crystal Ball functions (one
with a tail towards lower and one with a tail towards higher masses) whose parameters are
taken from simulation. The widths and the $B^0$ peak position are free to vary
in the fit while the $B^0_s$ peak offset is constrained in the same way as that
of the $B^0_s \!\rightarrow D^+D^-$ component. The combinatorial background is parametrized with
an exponential function, with separate exponents used for the final
states with two or three kaons. Partially reconstructed \mbox{$B^0 \!\rightarrow D^{*+}D^-$} decays with \mbox{$D^{*+} \!\rightarrow D^+\pi^0$}, where the
neutral pion is missed, lie outside the mass range used for the fit. The
equivalent \mbox{$B^0_s \!\rightarrow D^{*+}D^-$} decays and decay modes with only one or no charm meson,
such as \mbox{$B^0 \!\rightarrow D^-K^-K^+\pi^+$}, are also neglected in the mass fit. The
influence of their omission on the $C\!P$ measurement is treated as a systematic uncertainty.
The mass distribution is shown in \cref{fig:mass_and_decaytime} (a). The combined
\mbox{$B^0 \!\rightarrow D^+ D^-$} signal yield is \num{1610\pm50}, of which \num{1347\pm45} are in the Cabibbo-favored final state with two \mbox{$D^+ \!\rightarrow K^-\pi^+\pi^+$} decays.

The measurement of decay-time-dependent $C\!P$ violation requires knowledge of
the initial flavor of each reconstructed $B^0$ meson. Flavor-tagging
algorithms deliver a measured tag decision $d'$ for the flavor of the $B^0$
meson, which takes the value $+1$ for a $B^0$, $-1$ for a ${\kern 0.18em\overline{\kern -0.18em B}{}^0}$ initial
state, and $0$ if no decision is possible, and an estimate $\eta$ of the
probability for the tag decision to be incorrect. The latter is referred to as
the mistag probability. Two classes of flavor-tagging algorithms are used:
opposite-side (OS) and same-side (SS)
taggers~\cite{LHCb-PAPER-2011-027,LHCb-PAPER-2015-027,LHCb-PAPER-2016-SSTagger}.
In $b\overline{b}$ pair production, the dominant source of $b$ hadrons at LHCb, the
signal $B^0$ meson is accompanied by a second $b$ hadron. The OS taggers
determine the flavor of the signal by examining the decay
products of this second $b$ hadron. The information from the decay products
consists of the charge of muons or electrons produced in semileptonic decays,
the charge of kaons from $b \rightarrow c \rightarrow s$ transitions, the
charge of charm hadrons from $b \rightarrow c$ transitions, and the net
charge of all decay products. The SS taggers analyze pions and protons related
to the hadronization process of the $B^0$ meson. This is the first analysis to
use the LHCb SS proton and OS charm taggers, and the first to use the new SS
pion tagger.

The outputs of all OS algorithms are combined into an
overall OS tagging decision and mistag estimate, and the same is done for the SS
algorithms. The mistag
estimates $\eta \in \{\eta_{\text{OS}}, \eta_{\text{SS}}\}$ are calibrated using linear
functions $\omega(\eta | d)$, so that $\eta$ on average matches the true mistag
probability $\omega$, which depends on the true production flavor $d$ of the $B^0$
meson.
The calibration studies are performed with a sample of 
\mbox{$B^0 \!\rightarrow D^+_s D^-$} decays, for which the final state determines the flavor of the $B^0$ at decay.
Since the calibration and signal channels are
kinematically very similar, the calibration can be applied to the signal channel without further corrections.
To ensure that the same calibration is valid for both, the
same selection is used as for the signal decay with one \mbox{$D^+ \!\rightarrow K^-K^+\pi^+$},
apart from requiring that the $K^-K^+\pi^+$ invariant mass lie within
\SI[per-mode=symbol]{25}{\mega\eV\!\per\square\clight} of the known $D^+_s$ mass~\cite{PDG2014} and dropping the
vetoes against misidentified backgrounds. Background is subtracted from the calibration sample via the \mbox{\em sPlot} technique~\cite{Pivk:2004ty}. The tagging calibration parameters are
determined from a fit to the decay time and tag distributions of
\mbox{$B^0 \!\rightarrow D^+_s D^-$} candidates, in which the detection asymmetry, the production asymmetry of the $B^0$ mesons,
and the flavor-specific semileptonic asymmetry $a_{\text{sl}}^d$ are taken into
account. Here, the detection asymmetry describes the difference in reconstruction efficiency between the
$D^+_s D^-$ and $D^-_sD^+$ final states, and ${\ensuremath{A}_{\mathrm{P}}} \equiv {[\sigma(\kern 0.18em\overline{\kern -0.18em B}{}^0)-\sigma(B^0)]}/{[\sigma(\kern 0.18em\overline{\kern -0.18em B}{}^0)+\sigma(B^0)]}$, where
$\sigma$ denotes the production cross-section inside the LHCb acceptance.
The values of all these parameters are fixed according to the
latest LHCb measurements~\cite{LHCb-PAPER-2014-053,LHCb-PAPER-2014-042},
and their uncertainties are treated as sources of systematic uncertainty on the
calibration parameters. Further systematic uncertainties are assigned due
to the calibration method, the dependence of the efficiency on decay time, the decay time resolution, and the
background subtraction. More details on the calibration studies are given in \hyperref[app:tagging]{the supplemental material}.

In the \mbox{$B^0 \!\rightarrow D^+ D^-$} signal data sample, the correlation between the OS and the SS
mistag estimates is found to be negligible.
A small correlation of the mistag probability with decay time is seen; this is neglected in the
main fit but considered as a source of systematic uncertainty.

The effective tagging efficiency is the product of the probability for reaching
a tagging decision, $\varepsilon_{\mathrm{tag}}$ = \SI{87.6\pm0.8}{\percent}, and the square of the
effective dilution $D = 1 - 2\omega = \SI{30.3\pm1.1}{\percent}$. Its value is
$\varepsilon_{\mathrm{tag}} D^2= \SI{8.1\pm0.6}{\percent}$, the highest effective tagging efficiency
to date in tagged $C\!P$ violation measurements at LHCb thanks to the improved
flavor-tagging algorithms and the kinematic properties of the selected \mbox{$B^0 \!\rightarrow D^+ D^-$}
decays.


The $C\!P$ violation observables $S$ and $C$ are determined from a multidimensional
fit to the background-subtracted tag and decay time distributions of the tagged \mbox{$B^0 \!\rightarrow D^+ D^-$}
candidates; a projection of the decay time distribution summed over the non-zero tag decisions is shown in \cref{fig:mass_and_decaytime} (b).
The conditional PDF describing the reconstructed decay time $t'$ and tag
decisions $\vec{d'} = (d^{\prime}_{\text{OS}}, d^{\prime}_{\text{SS}})$, given a per-event decay time resolution
$\sigma_{t'}$ and per-event mistag probability estimates $\vec{\eta} = (\eta_{\text{OS}},
\eta_{\text{SS}})$, is
\begin{equation}\label{eq:fullpdf}
  P\left(t',\vec{d'}\:\vert\: \sigma_{t'},\vec{\eta}\right)
  \propto \epsilon(t') \left(\mathcal{P}(t,\vec{d'}\:\vert\: \vec{\eta})
    \otimes \mathcal{R}(t'-t\:\vert\: \sigma_{t'})\right)\,,
\end{equation}
where
\begin{equation}
  \mathcal{P}(t,\vect{d'}\given \vect{\eta}) \\
  \propto \sum_{d} \mathcal{P}(\vect{d'} \given d,\vect{\eta})
      [1 - d\, A_\text{P}] \,
      e^{-t/\tau}\left\{1 - d\, S \sin(\dm t) + d\, C \cos(\dm t)\right\}\,,
\end{equation}
and where $t$ is the true decay time, $d$ is the true production flavor,
$A_\text{P}$ is the production asymmetry, and $\mathcal{P}(\vec{d'} \:\vert\:
d,\vec{\eta})$ is a two-dimensional binomial PDF describing the distribution
of tagging decisions given $\vec{\eta}$ and $d$. Normalization factors are omitted for brevity. In the fit, the mass
difference $\mathrm{\Delta} m$ and the lifetime $\tau$ are constrained to their known values
within uncertainties~\cite{PDG2014}. The production asymmetry $A_{\text{P}}$ is
constrained separately for the \SIlist{7;8}{\tera\eV} samples to the values obtained from weighting the results from the
measurements in Ref.~\cite{LHCb-PAPER-2014-042} according to the kinematic
distribution of the $B^0$ signal candidates. The decay time resolution
model $\mathcal{R}$ is the sum of three Gaussian functions, two of which have
event-dependent widths proportional to $\sigma_{t'}$, and one which has a global
width that describes the effect of candidates matched to a wrong PV; all
three share a common mean. All parameters of the resolution model are
determined from simulation. The average decay time resolution in data is \SI{49}{\fs}. The function $\epsilon(t')$ describes the
efficiency for all reconstruction and selection steps as a function of
the reconstructed decay time. It is represented by cubic
splines~\cite{splines}, with the spline coefficients left unconstrained in the fit.

\begin{figure}[bt]
\centering
\includegraphics[width=0.48\textwidth]{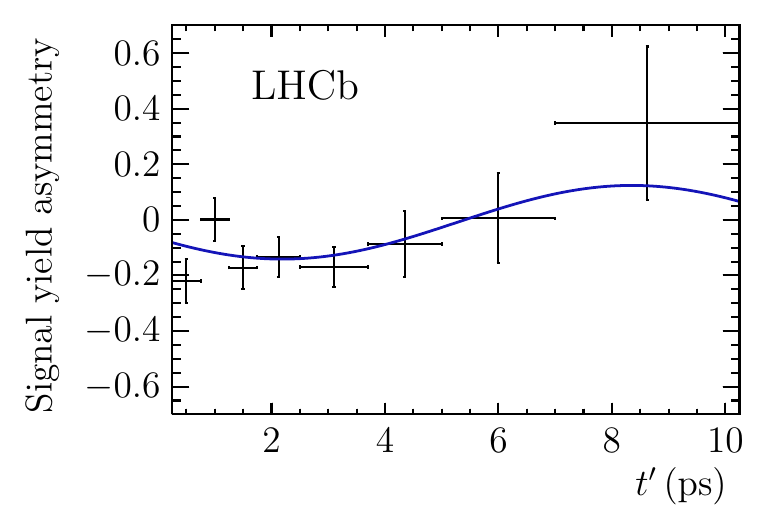}
\caption{Decay-time-dependent signal yield asymmetry. The solid curve is the
projection of the signal PDF given in \cref{eq:fullpdf}.}
\label{fig:asymmetry}
\end{figure}

The statistical uncertainties are estimated using the bootstrap
method~\cite{efron1979}. Individual bootstrap samples are
drawn from the candidates in data that pass the full selection; the analysis
procedure described above, consisting of the mass fit, background subtraction,
and decay time fit, is then applied to obtain the values of the $C\!P$
observables for each such sample. Two-sided
\SI{68}{\percent} confidence intervals, with equal tail probabilities on either side, are obtained from the distributions of
fitted parameters in the bootstrapped samples. To account for the uncertainties of
the flavor-tagging calibration parameters, which are fixed in the likelihood
fit, further pseudoexperiments are generated in which these
flavor-tagging calibration parameters are varied within their combined
statistical and systematic uncertainties. The results are then used to correct
the uncertainties from the bootstrapping procedure. The $C\!P$ observables are
measured to be $S = \num{-0.54}\,^{+0.17}_{-0.16}$ and $C =
\num{0.26}\,^{+0.18}_{-0.17}$ with a correlation coefficient of $\rho =
\num{0.48}$. The decay-time-dependent signal yield
asymmetry $(N_{{\kern 0.18em\overline{\kern -0.18em B}{}^0}} - N_{B^0})/(N_{{\kern 0.18em\overline{\kern -0.18em B}{}^0}} + N_{B^0})$, where $N_{B^0}$ is the number of $B^0 \!\rightarrow D^+ D^-$ decays with
a $B^0$ flavor tag, and $N_{{\kern 0.18em\overline{\kern -0.18em B}{}^0}}$ the number with a ${\kern 0.18em\overline{\kern -0.18em B}{}^0}$ tag, is shown in \cref{fig:asymmetry}.

Several sources of systematic uncertainties on the $C\!P$ observables are
studied with pseudoexperiments. The largest systematic uncertainty arises
from neglecting backgrounds in which the final state contains only one charm
meson, such as \mbox{$B^0 \!\rightarrow D^-K^-K^+\pi^+$}. The yield of these backgrounds is estimated to be
about \SI{2}{\percent} of the signal yield and their impact is assessed by
assuming that they maximally violate $C\!P$ symmetry and have the eigenvalue
opposite to the signal mode. This leads to a systematic uncertainty of
\num{\pm0.05} on $S$ and \num{\pm0.013} on $C$. Further systematic
uncertainties on $S$ are related to the assumption $\mathrm{\Delta}\Gamma = 0$
(\num{\pm0.014}), and to the modeling of the dependence of the efficiency on decay time
(\num{\pm0.007}). For $C$ the second largest systematic uncertainty of
\num{\pm0.007} is due to neglecting the correlation between the invariant mass
and the decay time. Additional systematic uncertainties arise from the decay time
resolution, the uncertainty on the knowledge of the length scale, the
parametrization of the mass model, and from uncertainties on the $B^0$
production asymmetry and mass difference $\mathrm{\Delta} m$. The total systematic
uncertainty, calculated as the sum in quadrature of all contributions, is
\num{\pm0.05} for $S$ and \num{\pm0.02} for $C$, with a correlation coefficient of $\rho = \num{-0.69}$.


In conclusion, a measurement of the $C\!P$ observables $S$ and $C$ in the
decay channel \mbox{$B^0 \!\rightarrow D^+ D^-$} is performed. Using the full data sample collected
by the LHCb experiment during Run~1, which corresponds to a total integrated
luminosity of \SI{3}{\per\femto\barn}, they are determined to be
\begin{align*}
  S &=  -0.54 \, ^{+0.17}_{-0.16} \, \text{(stat)} \pm 0.05 \, \text{(syst)}\,, \\
  C &=  \phantom{-}0.26 \, ^{+0.18}_{-0.17} \, \text{(stat)} \pm 0.02 \, \text{(syst)}\,,
\end{align*}
with a statistical correlation coefficient of $\rho = \num{0.48}$. This
result excludes the conservation of $C\!P$ symmetry by \num{4.0} standard deviations. It is compatible with the previous measurement by the BaBar
experiment of $S = \num[parse-numbers=false]{-0.63\pm0.36\pm0.05}$ and $C = \num[parse-numbers=false]{-0.07\pm0.23\pm0.03}$~\cite{Aubert:2008ah} while being significantly more precise. A
proper evaluation of the compatibility with the result from the Belle
experiment~\cite{Rohrken:2012ta} could not be performed due to its non-Gaussian uncertainties. The result presented here
corresponds to $\sin(\phi_d +\Delta \phi) = 0.56\,^{+0.16}_{-0.17}$, which
constrains the phase shift to the world's most precise value of \mbox{$\Delta\phi =
-0.16\,^{+0.19}_{-0.21}$\,\si{\radian}}, and thus implies only a small
contribution from higher-order Standard Model corrections.

\section*{Acknowledgements}

\noindent We express our gratitude to our colleagues in the CERN
accelerator departments for the excellent performance of the LHC. We
thank the technical and administrative staff at the LHCb
institutes. We acknowledge support from CERN and from the national
agencies: CAPES, CNPq, FAPERJ and FINEP (Brazil); NSFC (China);
CNRS/IN2P3 (France); BMBF, DFG and MPG (Germany); INFN (Italy); 
FOM and NWO (The Netherlands); MNiSW and NCN (Poland); MEN/IFA (Romania); 
MinES and FASO (Russia); MinECo (Spain); SNSF and SER (Switzerland); 
NASU (Ukraine); STFC (United Kingdom); NSF (USA).
We acknowledge the computing resources that are provided by CERN, IN2P3 (France), KIT and DESY (Germany), INFN (Italy), SURF (The Netherlands), PIC (Spain), GridPP (United Kingdom), RRCKI and Yandex LLC (Russia), CSCS (Switzerland), IFIN-HH (Romania), CBPF (Brazil), PL-GRID (Poland) and OSC (USA). We are indebted to the communities behind the multiple open 
source software packages on which we depend.
Individual groups or members have received support from AvH Foundation (Germany),
EPLANET, Marie Sk\l{}odowska-Curie Actions and ERC (European Union), 
Conseil G\'{e}n\'{e}ral de Haute-Savoie, Labex ENIGMASS and OCEVU, 
R\'{e}gion Auvergne (France), RFBR and Yandex LLC (Russia), GVA, XuntaGal and GENCAT (Spain), Herchel Smith Fund, The Royal Society, Royal Commission for the Exhibition of 1851 and the Leverhulme Trust (United Kingdom).

\addcontentsline{toc}{section}{References}
\setboolean{inbibliography}{true}
\bibliographystyle{LHCb}
\bibliography{main,LHCb-PAPER,LHCb-CONF,LHCb-DP,LHCb-TDR,external}

\newpage

\section*{Supplemental material}

\appendix
\subsection*{Tagging calibration}
\label{app:tagging}

\begin{figure}[!htb]
\centering
\includegraphics[width=0.48\textwidth]{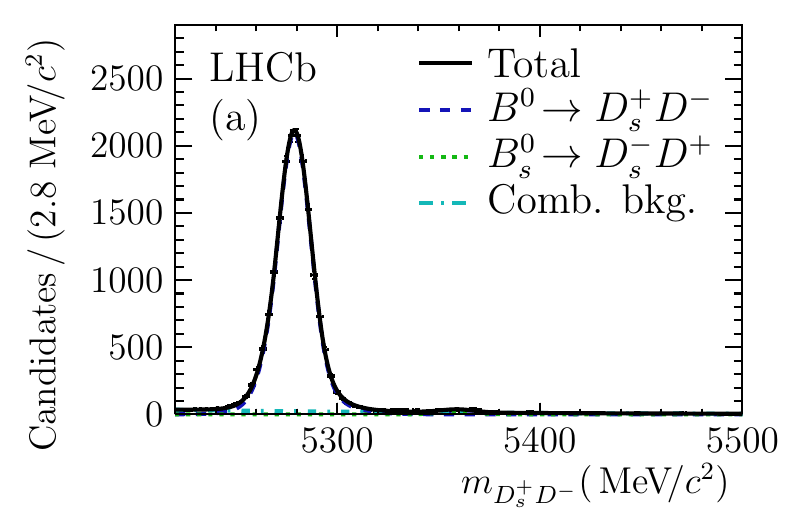}
\includegraphics[width=0.48\textwidth]{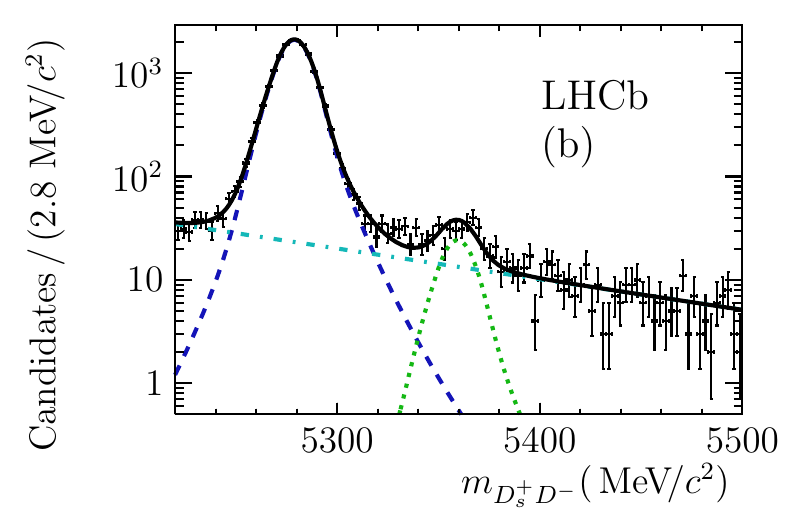}
\caption{Masses of \mbox{$B^0 \!\rightarrow D^+_s D^-$} candidates and
projected PDFs, shown with a linear scale on the vertical axis (a) and a logarithmic scale (b). The solid line is the PDF projection, the
blue dashed line represents the \mbox{$B^0 \!\rightarrow D^+_s D^-$} component, while the dash-dotted cyan
(dotted green) line represents the combinatorial ($B^0_s \!\rightarrow D^-_s D^+$) background.}
\label{fig:massDsD}
\end{figure}

For the flavor tagging calibration, a fit to the background-subtracted decay time distribution of \mbox{$B^0 \!\rightarrow D^+_s D^-$} candidates is used.
In the first step, the mass fit, the \mbox{$B^0 \!\rightarrow D^+_s D^-$} signal is modeled as two Crystal Ball functions that
share a common mean, but have different widths and tail parameters that are obtained from
simulations. The \mbox{$B^0_s \!\rightarrow D^-_s D^+$} background component is modeled similarly and shares all shape parameters with \mbox{$B^0 \!\rightarrow D^+_s D^-$}
except for the peak position, which is constrained to be $\SI[per-mode=symbol]{87.35}{\mega\eV\!\per\square\clight}$
larger. The combinatorial background component is modeled as an exponential
function. The \mbox{$B^0 \!\rightarrow D^+_s D^-$} yield is found to be $\num{16736\pm134}$. The invariant mass
distribution of \mbox{$B^0 \!\rightarrow D^+_s D^-$} candidates
is shown in \cref{fig:massDsD} with the PDF projection overlaid.

The decay time fit to the \mbox{$B^0 \!\rightarrow D^+_s D^-$} candidates uses a modified version of the
PDF from the \mbox{$B^0 \!\rightarrow D^+ D^-$} fit where $S$ and
$C$ are fixed to zero and unity, respectively, and the production flavor
variables $d^\prime$ and $d$ are replaced by the mixing state ($+1$ if the
production flavor and the reconstructed decay flavor are the same, and $-1$
otherwise).
Additional modifications are implemented to treat the production asymmetry
correctly after replacing $d'$ by the mixing state, to allow for a
flavor-specific asymmetry $a_{\mathrm{sl}}^d$ (fixed to the latest LHCb
measurement~\cite{LHCb-PAPER-2014-053}) and to include an asymmetry in the
detection efficiency for $D^+_sD^-$ and $D^-_sD^+$ (only for the evaluation of systematic uncertainties).
The $B^0$ oscillation frequency, $\mathrm{\Delta} m$, and the mean lifetime, $\tau$, are fixed.
The associated systematic uncertainties are taken to be the changes in the
calibration parameters when the quantities that were fixed are varied within
their uncertainties.

\begin{figure}[!htb]
\centering
\includegraphics[width=0.48\textwidth]{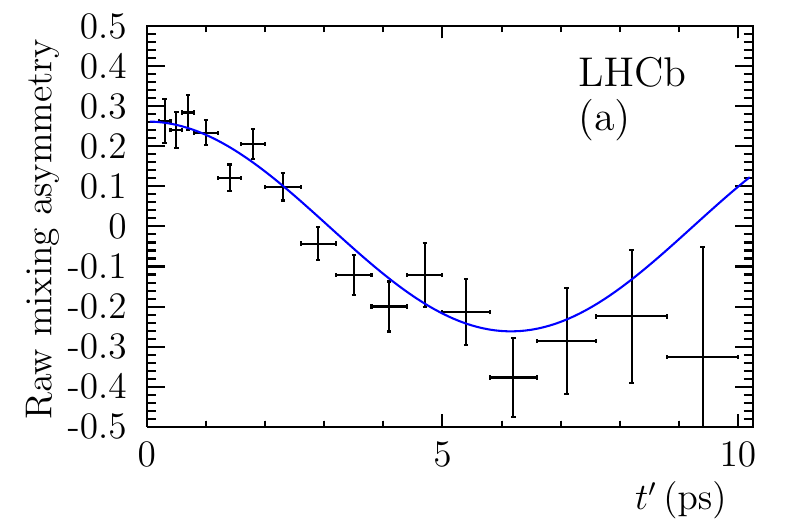}
\includegraphics[width=0.48\textwidth]{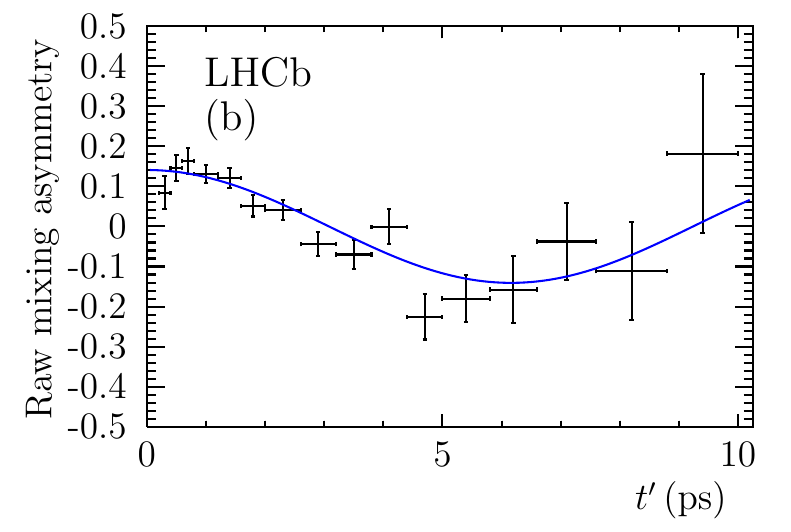}
\caption{Raw mixing asymmetry as a function of the $B^0$ decay time for events
tagged by (a) the OS tagger and (b) the SS tagger. The solid line
represents the PDF projection.}
\label{fig:asymDsD}
\end{figure}

The calibration function for initial $B^0$ and $\kern 0.18em\overline{\kern -0.18em B}{}^0$ mesons is given by
\begin{equation}
\omega(\eta \vert d) = p_0 + d\frac{\Delta p_0}{2} + \left(p_1 + d\frac{\Delta p_1}{2}\right)(\eta-\langle\eta\rangle)\,.
\end{equation}
Here, $d$ is $+1$ for mesons whose initial flavor is $B^0$ and $-1$ for
${\kern 0.18em\overline{\kern -0.18em B}{}^0}$. The predicted, per-candidate
mistag rate is $\eta$ and $\omega$ the calibrated, per-candidate mistag rate.
The value of $\eta$ when averaged over all tagged candidates is $\langle\eta\rangle$.
The calibration parameters are $p_0$, $\Delta p_0$, $p_1$, and $\Delta p_1$,
and are to be understood as follows. Averaging the calibrated mistag rate over
all candidates, $p_0$ is the value obtained when further averaging over
$B^0$ and $\Bdb$, and $\Delta p_0$ is the difference between the average
mistag rates for $B^0$ and $\Bdb$. The coefficients $p_1$ and $\Delta p_1$
describe the linear relationship between $\eta$ and $\omega$; again, $p_1$ is
averaged over $B^0$ and $\Bdb$ and $\Delta p_1$ is the difference between
$B^0$ and $\Bdb$.
For perfectly calibrated taggers the following two relations hold
\begin{equation}
  \begin{aligned}
    p_0 &= \langle \eta \rangle\\
    p_1 &= 1\,,
  \end{aligned}
\end{equation}
and the tagging asymmetries vanish: $\Delta p_1 = \Delta p_0 = 0$.
The OS calibration parameters are determined to be
\begin{equation}
  \begin{aligned}
p_{1,\mathrm{OS}} &= 1.07 \pm 0.07\, (\mathrm{stat}) \pm 0.01\, (\mathrm{syst})\,,\\
p_{0,\mathrm{OS}}  &= 0.369 \pm 0.008\, (\mathrm{stat}) \pm 0.010\, (\mathrm{syst})\,,\\
\langle\eta_{\mathrm{OS}}\rangle&= 0.3627\,,\\
\Delta p_{1,\mathrm{OS}} &= 0.03 \pm 0.11\, (\mathrm{stat}) \pm 0.03\, (\mathrm{syst})\,,\\
\Delta p_{0,\mathrm{OS}} &= -0.009 \pm 0.012\, (\mathrm{stat}) \pm 0.001\, (\mathrm{syst})\,.
  \end{aligned}
\end{equation}
The SS calibration parameters are determined to be
\begin{equation}
  \begin{aligned}
p_{1,\mathrm{SS}}  &= 0.84 \pm 0.09\, (\mathrm{stat}) \pm 0.01\, (\mathrm{syst})\,,\\
p_{0,\mathrm{SS}}  &= 0.430 \pm 0.006\, (\mathrm{stat}) \pm 0.009\, (\mathrm{syst})\,,\\
\langle\eta_{\mathrm{SS}} \rangle&= 0.4282\,,\\
\Delta p_{1,\mathrm{SS}} &= 0.07 \pm 0.13\, (\mathrm{stat}) \pm 0.05\, (\mathrm{syst})\,,\\
\Delta p_{0,\mathrm{SS}} &= -0.007 \pm 0.009\, (\mathrm{stat}) \pm 0.001\, (\mathrm{syst})\,.
  \end{aligned}
\end{equation}
The time-dependent raw mixing asymmetry $(N_{\text{unmixed}} -
N_{\text{mixed}})/(N_{\text{unmixed}} + N_{\text{mixed}})$, where
$N_{\text{unmixed}}$ is the number of \mbox{$B^0 \!\rightarrow D^+_s D^-$} decays with a final state that
does correspond to the flavor tag, and $N_{\text{mixed}}$ the number with a final state that does not,
as measured using OS or SS taggers, is shown in \cref{fig:asymDsD}.
In the \mbox{$B^0 \!\rightarrow D^+ D^-$} dataset, the tagging power for events which are tagged only by OS
taggers is \SI{1.02\pm0.09}{\percent}, and for events tagged only by SS
taggers \SI{1.36\pm0.19}{\percent}; for events tagged by both OS and SS taggers,
the combined tagging power is
\SI{5.7\pm0.5}{\percent}. These sum to a tagging power of \SI{8.1\pm0.5}{\percent} for events tagged by OS, SS, or OS and SS taggers.

\clearpage

\newpage
\centerline{\large\bf LHCb collaboration}
\begin{flushleft}
\small
R.~Aaij$^{40}$,
B.~Adeva$^{39}$,
M.~Adinolfi$^{48}$,
Z.~Ajaltouni$^{5}$,
S.~Akar$^{6}$,
J.~Albrecht$^{10}$,
F.~Alessio$^{40}$,
M.~Alexander$^{53}$,
S.~Ali$^{43}$,
G.~Alkhazov$^{31}$
% ...

{\footnotesize \it
% ...
$ ^{5}$Clermont Universit{\'e}, Universit{\'e} Blaise Pascal, CNRS/IN2P3, LPC, Clermont-Ferrand, France\\
$ ^{6}$CPPM, Aix-Marseille Universit{\'e}, CNRS/IN2P3, Marseille, France\\
% ...
$ ^{10}$Fakult{\"a}t Physik, Technische Universit{\"a}t Dortmund, Dortmund, Germany\\
% ...
$ ^{31}$Petersburg Nuclear Physics Institute (PNPI), Gatchina, Russia\\
% ...
$ ^{39}$Universidad de Santiago de Compostela, Santiago de Compostela, Spain\\
$ ^{40}$European Organization for Nuclear Research (CERN), Geneva, Switzerland\\
% ...
$ ^{43}$Nikhef National Institute for Subatomic Physics, Amsterdam, The Netherlands\\
% ...
$ ^{48}$H.H. Wills Physics Laboratory, University of Bristol, Bristol, United Kingdom\\
% ...
$ ^{53}$School of Physics and Astronomy, University of Glasgow, Glasgow, United Kingdom\\
% ...
}
\end{flushleft}

\end{document}